\documentclass[runningheads]{svmult}

\usepackage{makeidx}   
\usepackage{graphicx}  
\usepackage{subeqnar}  
\usepackage{multicol}  
\usepackage{physprbb}  
\makeindex             

\newcommand\kms{\rm ~km~s^{-1}}
\newcommand\Msun{M_\odot}
\newcommand\ml{M_\odot~\rm yr^{-1}}
\newcommand\gsim{\!\!\!\phantom{\ge}\smash{\buildrel{}\over
  {\lower2.5dd\hbox{$\buildrel{\lower2dd\hbox{$\displaystyle>$}}\over
                               \sim$}}}\,\,}

\begin{document}
\title*{Circumstellar Interaction Around Supernovae}
%
%
%
%
\titlerunning{Circumstellar Interaction Around Supernovae}
%
\author{Roger A. Chevalier}
\authorrunning{Chevalier}
%
%
\institute{Department of Astronomy, University of Virginia, P.O. Box 3818, \\
Charlottesville, VA 22903, USA
}

\maketitle              

\begin{abstract}
Circumstellar interaction has been observed 
around all types of massive star supernovae, especially at radio
and X-ray wavelengths.
The interaction shells in Type Ib/c supernovae appear to be moving
rapidly, although SN 1998bw remains the only case for which the
evidence points to relativistic expansion.
Type IIP supernovae have relatively low mass loss rates, consistent
with moderate mass single stars at the ends of their lives.
Type IIb and some IIL and IIn supernovae can have strong mass loss
leading up to the explosion, which occurs in a star with little
H on it.
The mass loss may be driven by the luminosity of the late burning stages.

\end{abstract}

\section{Introduction}

During their lifetimes, massive stars undergo mass loss during various
evolutionary phases.
When they finally end their lives as supernovae, the interaction
reveals the mass loss processes leading up to the explosion.
and the structure of the exploding star.
Recent multiwavelength observations have revealed many facets of
the interaction and
we now have  observations showing
 circumstellar interaction around all types
of massive star supernovae.
Circumstellar interaction can be related to late massive star evolution
and to the physical processes in the dense gas interaction.
Analysis of the radio and X-ray emission from supernovae
yields the energy in high velocity ejecta and can be used to set limits
on the amount of relativistic ejecta; the relation of supernovae
to gamma-ray bursts can thus be explored.

A recent review of circumstellar interaction with an emphasis on the
relevant physical processes is in \cite{CF02}.
Here, I briefly review recent developments, emphasizing the relation
of circumstellar interaction to the various supernova types.
In \S~2, I discuss the wide range of 
multiwavelength observations of circumstellar
interaction that is now available.
Interaction around the various types of supernovae is discussed in \S~3.
Conclusions are in \S~4.

\section{Multiwavelength observations of  interaction}

One of the most sensitive ways of finding or observing  supernovae
with circumstellar interaction
is through their nonthermal, radio synchrotron emission.
The radio light curves of extragalactic supernovae show an early absorbed
phase, so that there is a rise to a maximum followed by an approximately
power law decline \cite{W02}.
Fig. 1 shows the relationship between the observed peak luminosity and the
age at peak luminosity (from \cite{C98}, with updates).
Recent detections are SN 2001ig \cite{Ry01}, SN 2001gd \cite{St02}, 
and SN 2002ap \cite{B02}.
All types of massive star supernovae are represented, including
Type Ib/c which are thought to have lost their hydrogen envelopes
before the explosion, Type IIb which have lost almost all
their hydrogen envelopes
before the explosion, Type IIn with narrow line optical emission,
Type IIL with linear optical light curves, Type IIP with plateau
light curves, and the peculiar SN 1987A.
Interestingly, the different types of supernovae populate different
parts of this diagram.   Although the numbers are still small, some
trends are showing up in the interaction properties.
The Type Ib/c supernovae peak early and have a range of luminosity,
while the Type IIL and Type IIn SNe peak late and also have a range
of luminosity.
The Type IIb are intermediate between these and the one Type IIP
is of low luminosity.
The early peak of the Type Ib/c supernovae is an indication that
their radio emitting regions expand relatively rapidly \cite{C98}.

\begin{figure}[t]
\centering
\includegraphics[viewport=  0 0 600 600,scale=0.40,angle=0,origin=c,clip]
{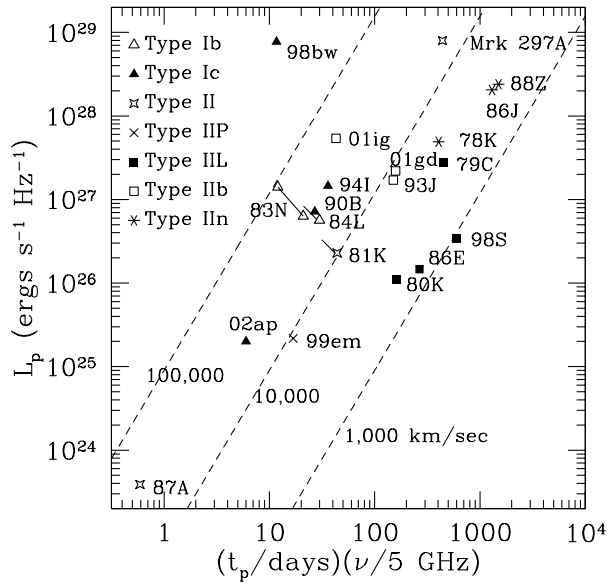}
\caption[]{ Peak luminosity and corresponding epoch for the
well-observed radio supernovae. The dashed lines give curves of constant
expansion velocity, {\it assuming} synchrotron self-absorption (see \cite{C98}).
}
\label{fig1}
\end{figure}

The other wavelength region where young supernova remnants are
broadly observed is X-rays.
Fig. 2 shows the relation between the X-ray and radio luminosity of young
remnants.
Here, the radio luminosity is estimated from $\nu L_{\nu}$ at 
5 GHz.  
X-ray emission has been detected from the full range of massive star
supernovae, and there is a clear correlation of the X-ray and radio
emission.
The X-ray luminosities, primarily from \cite{IL02}, are not all in the
same X-ray band, but this should not have a large effect on the correlation.
X-ray line emission has been detected in a number of cases 
(SN 1987A \cite{Mi02}, SN 1986J \cite{He98}, SN 1993J \cite{Sw02},
SN 1998S \cite{Pe02}, Cas A), showing that the emission is thermal, at least in these cases.
The  correlation of the X-ray and radio emission shows that both
of these give a measure of the strength of circumstellar interaction,
even though they involve different radiation mechanisms.
Radio emission provides a good indicator of whether a particular
event is suitable for observation with the space missions {\it Chandra}
and {\it XMM}.

\begin{figure}[t]
\centering
\includegraphics[viewport=  0 0 600 600,scale=0.40,angle=0,origin=c,clip]
{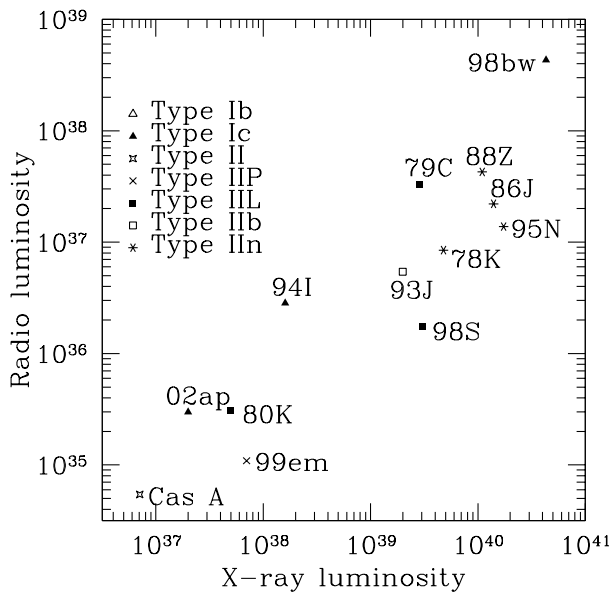}
\caption[]{Radio luminosity vs. X-ray luminosity of supernovae, both
in ergs s$^{-1}$.  The
X-ray data are primarily from \cite{IL02}.}
\label{fig2}
\end{figure}

At UVOIR (ultraviolet/optical/infrared) wavelengths, circumstellar interaction
is generally not observed in Type Ib/c and Type IIP supernovae, but
is observed in the spectra of Type IIL, IIb, and IIn supernovae.
The reason appears to be the relatively low densities and high velocities
present in Type Ib/c and Type IIP supernovae, so that radiative shock
waves are not present; in addition, the  X-ray luminosities are not
sufficient to
illuminate the expanding ejecta to observable levels.
In the cases with stronger circumstellar interaction, UVOIR emission
is potentially observable from shocked, cooled gas, from freely expanding
ejecta, and from unshocked circumstellar gas.
In the case of SN 1995N,  all of these components have
been observed \cite{Fe01};
ultraviolet {\it HST} observations are especially useful for observing
the ejecta.

Strong infrared emission, presumably from dust, has been observed
from the Type IIL supernovae SN 1979C and 1980K \cite{D83}
and, recently, from a number of Type IIn supernovae \cite{Fa00,Di94,Ge02}.
The preferred explanation has been emission from radiatively
heated circumstellar dust.
Another possible source of emission is collisionally heated dust
in the hot, shocked gas, as is observed in a number of Galactic
supernova remnants, including Cas A \cite{D87}.
The high observed luminosity implies that the source is not dust formed in
the freely expanding ejecta.

\section{Circumstellar interaction and supernova types}

\subsection{Type Ib/c supernovae and ``hypernovae''}

The best observed Type Ic supernova was SN 1994I in M51, which had
a rapid decline after maximum light and can be interpreted as an
explosion with an ejecta mass of only $0.9\Msun$ \cite{Iw94,Yo95}.
It is the only normal Type Ic supernova to have been detected in
X-rays \cite{I98,I02}.
Chevalier \cite{C98} applied a synchrotron self-absorption model to
the available radio data on SN 1994I and found an implied shock radius
of about $1.2\times 10^{16}$ cm at an age of 36 days,
giving a shock velocity of $39,000\kms$.
Application of the asymptotic power law
density profile of \cite{MM99}
($\rho\propto r^{-10.1}$) yields an outer shock radius 
$R=1.7 \times 10^{16} (M/{\Msun})^{-0.32}   
(E/10^{51}{\rm~ergs})^{0.44} (\dot M_{-5}/v_{w3})^{-0.12}$ cm,
where $M$ is the ejecta mass, $E$ is the explosion energy,
$\dot M_{-5}=\dot M/10^{-5}\ml$ is the mass loss rate, and $v_{w3}=v_w/10^3 
{\rm km~s^{-1}}$ is the wind velocity \cite{B02}.
The approximate  
agreement of the hydrodynamic model 
with the radio radius
may be improved by using a more accurate density model for the exploded
star, because the density profile is not fully in the asymptotic power
law regime for such a small mass star.
Immler et al. \cite{I02} have used {\it Chandra} observations of SN 1994I
to deduce a presupernova mass loss rate of $~1\times 10^{-5}{\ml} (v_w/10\kms)$.
For a Wolf-Rayet star wind velocity $v_w\approx 10^3\kms$, the inferred mass
loss rate of $~1\times 10^{-3}\ml$ is surprisingly high.
However, Immler et al. \cite{I02} used a model for the interaction from
Chevalier \cite{C84}, which was developed before the recognition of Type Ib/c
supernovae as a separate type and was based on a Type Ia interpretation.
If the high mass loss rate holds up in a more detailed model, it may
be evidence for mass loss from a companion star, as has previously
been suggested based on radio observations of Type Ib/c supernovae
\cite{vD93}.

Much attention has recently been focused on Type Ic supernovae because of
the occurrence of SN 1998bw, apparently associated with the gamma-ray burst
GRB 980425 \cite{Ge98}.
The strong radio emission from the source 
at early times (Fig. 1) implied the presence of
relativistic motions, further supporting the GRB connection \cite{K98,Li99}.  
SN 1998bw also had an unusual optical spectrum, with velocities in lines
 extending out to $\sim 60,000\kms$ \cite{Ge98}.
The interpretation of the supernova light curve and spectra led to
an explosion energy estimate of $(2-4)\times 10^{52}$ ergs, extraordinarily
high for a supernova \cite{Iw98,Wo99}.
Other Type Ic supernovae have been discovered with broad lines 
in their spectra, including
SN 1997ef, SN 2002ap, and SN 2002bl.
Because of the high energies inferred for these events,
 Mazzali et al. \cite{Ma02} and others have called these objects
hypernovae.

Outside of SN 1998bw,   broad-lined Type Ic supernovae have not been directly
connected with   GRBs.
In the case of SN 2002ap, there are limits on any corresponding GRB
\cite{Hu02}.
Despite its broad line spectrum, SN 2002ap was a weaker radio source than
the normal SN 1994I (Fig. 1; \cite{B02}).
Its position in Fig. 1 indicates a higher $M_{ej}$ or lower $E$ than
SN 1994I, plus a lower circumstellar density.
The model of Mazzali et al. \cite{Ma02} has $M_{ej}=2.4 \Msun$, which can account
for part of the difference, but there is no evidence for the high energy
($4\times 10^{51}$ ergs) that they infer for the explosion.
X-ray observations of SN 2002ap with {\it XMM} also showed a surprisingly
low luminosity \cite{Ro02}.
This emission is
sensitive to the distribution of high velocity gas and thus sets additional
constraints to those obtained from optical observations.
Sutaria et al. \cite{Su02} suggested that inverse Compton is the most
likely X-ray emission mechanism because of the relatively soft spectrum
of the source.
Approximate modeling of the radio emission in \cite{B02} showed that
the energy in high velocity ejecta is small.
The claim of high energies in broad line SN Ic is based on modeling
the broad optical absorption lines.
However, the radio emission probably gives a more direct estimate of
the energy in high velocity ejecta and the possible relation to GRBs.
The broad line Type Ic SN 2002bl was also a much weaker radio
source than SN 1998bw \cite{Be02}.

The combination of radio and X-ray observations with a hydrodynamic
model for the exploded star that is consistent with the observed optical
supernova emission should make it possible to estimate the mass
loss density around the supernova progenitor.
This information will be useful in determining the progenitor
evolution leading to an explosion.
Type Ib/c supernovae occur at higher rates than expected for single stars
that have lost their envelopes and binary evolution is a likely factor
(e.g., \cite{We99}).
The mass loss properties remain uncertain parameters for the presupernova
evolution of these objects.

\subsection{Type IIb supernovae}

The Type IIb supernova SN 1993J was in M81 at a distance of only
about 3.6 Mpc and has become the second best observed case of circumstellar
interaction, after SN 1987A.
The exploded star in this case is thought to be an initially
$12-15 ~\Msun$ star that lost most of its envelope mass as a result
of binary interaction \cite{No93,Wo94}.

VLBI imaging shows an approximately symmetric, but variable,
shell of radio emission \cite{B00}.
The ratio of shell thickness to radius is $\Delta R/R\sim 0.3$ \cite{M95},
which is larger than expected in standard interaction models \cite{C82a}.
A possible explanation is that inhomogenities in the wind lead to
a broader region.
Although this specific situation has not been modeled, 
there is evidence that shock interaction with a clumpy medium gives
rise to a broader, more turbulent shocked medium than would otherwise
be present \cite{JJN96}.
Relatively large scale inhomogeneities may be related to the variability
that has been observed in the radio structure \cite{B00}.
An examination of the optical line profiles of the supernova also gives
an indication of inhomogeneity.
Without inhomogeneities, the corrugation of the reverse shock wave
is relatively small \cite{CBE92}, which would
produce a more box-like H$\alpha$ line than was observed
\cite{M00b}.

There has been some confusion regarding the ambient density profile
for SN 1993J.
Assuming a form $\rho_w\propto r^{-s}$, both radio 
\cite{vD94,FLC96,M97,Mi01}
and X-ray \cite{I01} studies have led to $s=1.5-1.7$.
This is a surprising result because it implies a mass loss rate
decreasing as a power law in time leading up to the supernova;
Immler et al. \cite{I01} deduce a gradual reduction in $\dot M$ by $> 10$
based on the X-ray light curve from $\it ROSAT$ observations.
However, the initial radio studies assumed that free-free absorption
was the primary absorption process.
Fransson \& Bj\"ornsson \cite{FB98} showed that other physical processes,
especially synchrotron self-absorption, are important for the radio
curve and a detailed model yields $s=2$, which is the expected value
close to a massive star.
The X-ray result of Immler et al. \cite{I01} appears to be due to an interpretation
of the slowly decaying X-ray flux in terms of an adiabatic shocked region.
In fact, the soft X-ray emission in the $\it ROSAT$ band is dominated
by emission from the reverse shock, which we expect to be radiative 
\cite{FLC96}.
The X-ray spectra from {\it ASCA} give further evidence for
a cool shell formed downstream from a radiative shock \cite{Un02}.
In the radiative case with $s=2$, the X-ray emission is expected to be
relatively constant when it is not absorbed \cite{CF94}.
Swartz et al. \cite{Sw02} have obtained a {\it Chandra} spectrum of
SN 1993J at an age of 7 years.
It shows  cool and hot components that can be identified with the reverse
shock wave and the forward shock, respectively.
The cool component itself requires two temperature components (0.35
and 1.01 keV) to fit the spectrum.
This may be an indication that the emission is from a radiatively
cooling shock front.

\subsection{Type IIn/IIL supernovae}

The Type IIn and IIL designations are not necessarily distinct;
for example, SN 1998S had a light curve that led to a IIL designation
\cite{L00}, but also had narrow lines, leading to a IIn designation 
\cite{Le00}.
SN 1995N and SN 1998S are the best observed recent supernovae in the IIn category.
There have been recent {\it HST} \cite{Fe01} and
X-ray \cite{Fo00,Pe02} observations of these
supernovae.
They both have X-ray luminosities in the range $10^{40}-10^{41} {\rm~erg~s^{-1}}$,
are strong radio emitters, and show narrow lines in their optical
spectra.
Both supernovae show evidence for the reverse shock front moving
into heavy element rich gas.
In the case of SN 1995N, the UV/optical spectra show evidence for
emission from photoionized gas just inside the reverse shock front,
including the OI $\lambda$7774 recombination line, but there is
no H emission associated with this component \cite{Fe01}.
{\it Chandra} spectra for SN 1995N are not yet available, but
summed {\it Chandra} spectra of SN 1998S show evidence for an
overabundance of the Si group elements \cite{Pe02}.
Thus, although these supernovae were of Type II, with hydrogen in
their spectra at early times, they  developed evidence for
a reverse shock front moving into enriched material within several
years of the explosion.
In addition, Leonard et al. \cite{Le00} noted that the photospheric spectrum
of SN 1998S at an age of 25 days  resembled that of a Type Ic
supernova, with absorption lines of OI $\lambda$7774 and
SiII $\lambda$6355 and weak H$\alpha$ emission.
The implication is that strong mass loss just before the supernova
almost entirely removed the hydrogen envelope.

Additional observations of SN 1998S have shown a number of interesting
features.
The evolution of the optical spectrum has been interpreted as
indicating several regions of mass loss through which  the supernova shock
front is moving  \cite{Le00,Fa00}.
The supernova, which was discovered on 3 March, 1998, showed symmetric
broad lines with FWZI (full width at zero intensity) of $\sim 20,000\kms$
until 12 March, 1998.
The sudden disappearance of the lines can be interpreted as due to
the shock wave overtaking an inner dense region.
Chugai \cite{C01} has recently interpreted the broad  lines
 as due to scattering in a dense
circumstellar medium out to $\sim 10^{15}$ cm.
This requires $\dot M_{-5}/v_{w1}\gsim 300$ in this region.
Leonard et al. \cite{Le00} find little evidence for 
subsequent circumstellar interaction until an age of 108 days.
The later X-ray and radio observations imply $\dot M_{-5}/v_{w1}\sim (10-20)$
\cite{Pe02}.
An early infrared excess implies the presence of circumstellar
dust \cite{Fa00}, which places additional constraints on the
circumstellar density and dust content.

Fassia et al. \cite{Fa01} observed narrow emission lines and inferred
that the outer circumstellar medium had a velocity of
$40-50\kms$.
From the time evolution and velocities, they inferred that the outer wind
started at least 170 years ago, stopped about 20 years ago, and the
inner dense wind may have started less than 9 years ago.
The variability might be related to the mass loss variability observed
in late type stars.

The structure of red supergiant winds is  uncertain, because
they are generally quite distant.
However, red giant AGB (asymptotic giant branch) stars are known
to have a semi-periodic shell structure, observed in scattered
light (e.g., IRC +10216, \cite{MH00}) as well as in
the {\it HST} images of several planetary nebulae (e.g., \cite{B01}).
The characteristic time scale for the shells, several 100 years,
is indicative of an instability in the acceleration mechanism for
the wind.
Similar mass loss processes are likely to occur in red supergiants;
the mass loss characteristics of IRC +10216 are $v_w=14\kms$
and $\dot M=2\times 10^{-5}\ml$, which are comparable to values
for red supergiants.
The shell properties for IRC +10216 can be determined
from observations of dust scattered light \cite{MH00}.
The shells have a density contrast of a factor 3 over the
rest of the wind and contain a substantial fraction of the
total mass in the wind.
If the supernova is expanding at $\sim 10^4\kms$, the interaction should
lead to structure in the supernova light curves on a timescale of $\sim 100$ days.

It is also possible that the mass loss properties are determined by
the late nuclear burning phases just before the explosion.
When there is very little H envelope left on a star, the nuclear
processes can play a role in the mass loss properties.
Both SN 1995N and SN 1998S appear to have exploded at a time
of strong mass loss when the H envelope was almost entirely lost.
SN 1993J also exploded with a modest H envelope (few $0.1\Msun$);
in this case, the small envelope mass has been attributed to
binary interaction \cite{No93,Wo94}.
However, the mass loss from the envelope should then be especially
strong when the star first becomes a red supergiant and is
not very powerful at the time of the explosion.
SN 1993J in fact had strong mass loss at the time of the explosion
and in the  cases of SN 1995N and SN 1998S, the mass loss was even
stronger.
These observations suggest the possibility that the heightened luminosity
in the late burning stages plays a role in driving the strong mass
loss that leads to the loss of the H envelope.

\subsection{Type IIP supernovae}

There is little evidence for circumstellar interaction around Type IIP
supernovae, but the evidence that we have indicates a low circumstellar
density.
Fig. 1 shows the relatively low radio luminosity of SN 1999em.
The X-ray luminosity of SN 1999em is correspondingly low (Fig. 2;
\cite{Pe02}) and the X-ray luminosity of the Type IIP SN 1999gi
is similarly low \cite{Sc01}.
Approximate models for the X-ray emission lead to a mass loss rate
of $(1-2)\times 10^{-6}{\ml} (v_w/10\kms)$.

Information on the initial mass of supernovae comes from studies of
their stellar environments, especially with {\it HST}; such studies
have been carried out for 
 the Type IIP SNe 1999em and 1999gi \cite{Sme02}.
These studies indicate that these Type IIP supernovae may 
have come from $8-12~\Msun$
stars.
Type IIP supernovae are compatible with single star evolution and
initial masses $\sim 8-15\Msun$ and relatively little mass loss during
the evolution.
This is consistent with the evolutionary models of Schaller et al. \cite{Sc92}
who find that these stars end their lives as RSGs with
$\dot M \sim 2\times 10^{-6}\ml$, in agreement with the properties noted above.

\section{Conclusions}

Early circumstellar interaction gives rise to a number of interesting
physical processes in dense media, including radiative
$\sim 1,000\kms$ shock waves, particle acceleration in fast shock
waves,  and the emission
of infrared dust echoes.
The conditions in circumstellar interaction approach those in
quasars; in fact, some supernovae which are likely circumstellar
interactors have spectra similar to
Seyfert galaxies \cite{Fi89}, but the physical situation is better understood
for the supernova case.
Analysis of the observations shows the mass loss processes leading to
the supernova.  

The observational support for these studies is promising.
The {\it Chandra} and {\it XMM-Newton} X-ray missions have made
the X-ray emission from nearby supernovae accessible.
{\it SOFIA} and {\it SIRTF} should contribute to our understanding of infrared
dust emission from supernovae.  
Together with ground-based observations, the detailed picture of supernova
interaction with circumstellar matter should give considerable insight
into the late evolution of massive stars.

I am grateful to Claes Fransson for collaboration and discussion of
these topics.
This research was supported in part by NASA grant NAG 5-8232 and
grant GO-08648.03-A
from the Space Telescope Science Institute, which is operated by the
Association of Universities for Research in Astronomy, Inc., under NASA
contract NAS 5-26555.

%

\end{document}